\documentclass[pre,reprint,showpacs]{revtex4-1}                  
             
\usepackage{graphicx}
\usepackage{amssymb}
\usepackage{amsmath}
\usepackage{subfigure,wrapfig}

\begin{document}

\title{Interface growth in the channel geometry and tripolar Loewner evolutions}

\author{Miguel A.~Dur\'an} \email{mdr1angel@df.ufpe.br}
\author{Giovani L.~Vasconcelos}
\email{giovani@df.ufpe.br}

\affiliation{Laborat\'orio de F\'{\i}sica Te\'orica e Computacional, 
Departamento de F\'{\i}sica, Universidade Federal de Pernambuco,
50670-901, Recife, Brazil.}

\date{\today}

\begin{abstract}
A class of  Laplacian growth models  in the channel geometry is studied  using the formalism of  \emph{tripolar} Loewner evolutions, in which  three points, namely, the channel corners and infinity, are kept fixed.   Initially, the problem of fingered growth, where growth takes place only at the tips of slit-like fingers,  is revisited and a class of exact  exact solutions of the corresponding Loewner equation is presented for the case of stationary driving functions.  A model for interface growth is then formulated in terms of a generalized tripolar Loewner equation  and several examples are presented, including interfaces with multiple tips  as well as multiple growing interfaces. The model exhibits interesting dynamical features, such as  tip and finger competition.
\end{abstract}


\maketitle

\section{Introduction}

The Loewner equation describes a rather general class of growth processes in two dimensions where a curve  starts from a given point  on the boundary of a domain $\mathbb{P}$ in the complex $z$-plane  and grows into the interior of $\mathbb{P}$.  More specifically, the Loewner equation \cite{loewner} is a first-order differential equation for the conformal mapping $g_t(z)$ from the `physical domain,' consisting of the region  $\mathbb{P}$  minus the curve, onto a `mathematical domain' represented by $\mathbb{P}$ itself. The specific form of the Loewner equation depends both on  the domain $\mathbb{P}$ and on the point where the trace ends.  The most studied cases are the chordal and radial Loewner equations \cite{review3}, where in the former case the trace ends on a given point on the boundary, while in the latter it ends on a point  in the interior. Other geometries such as the dipolar case \cite{BB2005} and the channel geometry \cite{poloneses} have also been considered in the literature. The Loewner equation depends explicitly on a driving function, here denoted by  $a(t)$, which is the image of the growing tip under the mapping $g_t(z)$.  An important development in the field was the discovery by Schramm \cite{schramm} that when $a(t)$ is a Brownian motion the resulting Loewner evolution describes  the scaling limit of certain statistical mechanics models. This result spurred great interest in the so-called stochastic Loewner equation (SLE) and the subject  has now been widely reviewed, see, e.g. \cite{review1,review2,cardy,BB}.

Although the SLE has attracted most of the attention lately, the deterministic version of the Loewner equation  remains of considerable interest  both in a purely mathematical context \cite{kadanoff_jsp} and in connection with applications to growth processes \cite{selander,makarov,poloneses}  and integrable systems theory \cite{zabrodin2}. Indeed, the deterministic Loewner equation has been used to study Laplacian fingered-growth in both the chordal and radial cases \cite{selander,makarov} as well as in  the channel geometry \cite{poloneses}. In this class of models, growth takes place only at the tips of slit-like fingers and the driving function $a(t)$ has to follow a specific time evolution in order to ensure that the tip grows along gradient lines of the corresponding Laplacian field. Although this  thin-finger model \cite{poloneses} was able to reproduce some of the qualitative behavior seen in experiments  \cite{combustion},  treating the fingers as infinitesimally thin is a rather severe approximation.  

More recently, the growth of  ``extended fingers''  \cite{us_pre} or ``fat slits'' \cite{zabrodin1}, meaning a domain  encircled by an interface with endpoints on the real axis,  was considered within the formalism of deterministic Loewner evolutions  in the upper half-plane.  In particular, the interface growth model considered in Ref.~\cite{us_pre} was shown to exhibit 
interesting dynamical features akin to the finger competition mechanism observed in actual Laplacian growth \cite{pelce}.   Many  growth processes, however, take place in a bounded domain, e.g., within a channel \cite{pelce}, for which a Loewner-equation approach was still lacking. To develop a formalism based on Loewner evolutions to describe the  growth of an interface in the channel geometry is thus the main goal of the present study. Because the Loewner function $g_t(z)$ fixes the points $z=\pm 1$ (the channel corners) and  $z=i\infty$ (the channel `open' end),  we shall refer to growth processes in the channel geometry as  \emph{tripolar}  Loewner evolutions, to distinguish it from the dipolar case where two points are kept fixed \cite{BB2005}. 

We begin our analysis by revisiting  in Sec.~\ref{sec:2} the problem of fingered growth in the channel geometry, where a curve (finger) grows from a point on the real axis into a semi-infinite channel. We present a brief derivation of the corresponding tripolar Loewner and  report a new class of exact solutions  for the case of multiple fingers with stationary driving functions.  In Sec.~\ref{sec:3} we  then discuss a general class of models where a domain, bounded by an interface, grows into a channel starting from a segment on the real axis. In our model, the growth rate is specified at a certain number of special points on the interface, referred to as \emph{tips} and \emph{troughs}, which then determine the growth rate at the other points of the interface according to a specific growth rule (formulated in terms of a polygonal curve in the mathematical plane). A generalized tripolar Loewner equation for this  problem  is derived and several examples are given in which the interface evolution is obtained by a direct numerical integration of the Loewner equation.   
Our main results and conclusions are summarized in Sec.~\ref{sec:4}.

\section{Loewner Evolution for Fingers in the Channel}
\label{sec:2}

\subsection{Tripolar Loewner Equation}

\begin{figure}[t]
\begin{center}
\includegraphics[width=0.45\textwidth]{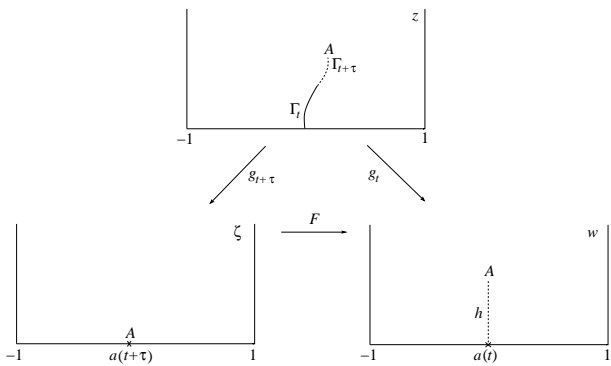}
\end{center}
\caption{The physical $z$-plane and the mathematical planes at times $t$ and $t+\tau$ for a single  curve in the channel geometry.   The Loewner function $g_t(z)$ maps the physical domain in the $z$-plane onto the channel in the mathematical $w$-plane, with the tip $\gamma(t)$ of the curve $\Gamma_t$  being mapped to the point $w=a(t)$. The portion (dashed line) of the curve accrued during a subsequent infinitesimal time interval $\tau$ is mapped under $g_t(z)$ to a vertical slit; see text.}
\label{fig:1}
\end{figure}

In order to set the stage for the remainder of the paper and to establish the relevant notation, we begin our discussion by considering  the simplest  Loewner evolution in the channel geometry, namely, that  in which a single curve grows from a point on the real axis into a semi-infinite channel whose side walls are placed at $x=\pm1$ and the bottom wall is at $y=0$; see Fig.~\ref{fig:1}. Here the relevant domain, $\mathbb{P}$, for the Loewner evolution  is
\begin{equation*}
\mathbb{P} = \left\lbrace {z= x+\mbox{i}y \in \mathbb{C}: y > 0, x \in ]-1,1[ } \right\rbrace .
\end{equation*}
The curve at time $t$ is denoted by $\Gamma_{t}$ and its growing tip is labeled by $\gamma(t)$. Now let  $w=g_t(z)$ be the conformal mapping that maps the  `physical domain,'  corresponding to the upper half-channel $\mathbb{P}$ in the $z$-plane  minus the curve $\Gamma_t$, onto the upper half-channel  $\mathbb{P}$ in an auxiliary  complex $w$-plane, referred to as the `mathematical plane,' i.e., we have 
  \begin{equation*}
  g_t: \mathbb{P} \backslash \Gamma_{t}  \rightarrow \mathbb{P}  ,
  \end{equation*}
with the curve tip $\gamma(t)$ being mapped to a point $w=a(t)$ on the real axis in the $w$-plane; see Fig.~\ref{fig:1}.

The mapping function $g_t(z)$ is required to satisfy  the so-called `hydrodynamic condition' at infinity, namely,
\begin{equation}
\label{eq:C}
 g_t(z) = -iC(t) + z+  O(\frac{1}{|z|}) , \qquad {{\rm Im}(z) \rightarrow \infty} ,
\end{equation}
where $C(t)$ is a real-valued, monotonously increasing function of time satisfying the condition $C(0)=0$. 
Condition (\ref{eq:C}), which implies $g_t'(i\infty)=1$, together with the requirement  $g_t(\pm1)=\pm1$, uniquely determines (up to a reparametrization of the time $t$) the mapping  $g_t(z)$.  The time parametrization is specified by fixing the function $C(t)$, which in turn is related to the growing rate of the curve; see below. 
The Loewner function $g_t(z)$ also satisfies the initial condition
\begin{equation}
g_{0}(z)=z,
\label{eq:f0}
\end{equation}
since we start with an empty channel.

We consider the growth process to be such that growth occurs only the tip of the curve and in such a way that the accrued portion from time $t$ to time $t+\tau$, where $\tau$ is an infinitesimal time interval, is mapped under $g_{t}(z)$ to a vertical slit of (infinitesimal) height $h$ in the mathematical $w$-plane; see Fig.~\ref{fig:1}.  For convenience of notation, we shall represent  the mathematical plane at time $t+\tau$ as the complex $\zeta$-plane, so that the mapping $\zeta=g_{t+\tau}(z)$ maps the physical domain at time $t+\tau$ onto the upper half-channel $\mathbb{P}$ in the $\zeta$-plane, with the new tip $\gamma(t+\tau)$ being mapped to the point $\zeta=a(t+\tau)$.

In order to derive the Loewner equation for the problem formulated above it is necessary to write $g_{t}$ in terms of $g_{t+\tau}$, i.e., $g_{t}=F(g_{t+\tau})$, where $w=F(\zeta)$ is the conformal mapping between the mathematical $\zeta$- and $w$- planes, and then consider the limit $\tau\to0$.  Alternatively, one can transform the  growth problem defined in the channel geometry into a corresponding Loewner evolution in the upper half-plane. To do that, we apply the following transformations:
\begin{equation}
\tilde{z}=\phi(z),\qquad 
\tilde{w}=\phi(w),
\qquad
 \tilde{\zeta}=\phi(\zeta), 
 \label{eq:sin}
\end{equation}
where
\begin{equation}
\phi(z)=\sin \left(\frac{\pi}{2}z\right),
\label{eq:sinz}
\end{equation}
so that  the channels in the
 $z$-,  $w$-, and $\zeta$- planes are mapped to the upper half-plane in the auxiliary complex $\tilde{z}$-, $\tilde{w}$-, and $\tilde{\zeta}$- planes, respectively.
 Let us also introduce introduce the following notation:
\begin{equation}
\label{eq:gtil}
\tilde{g}_{t}= \phi\circ g_t\circ\phi^{-1} .
\end{equation}
Next, consider the mapping $\tilde{w}=\tilde{F}(\tilde{\zeta})$ from the upper half-$\tilde{\zeta}$-plane onto the  upper half-$\tilde{w}$-plane with a vertical slit, so that one  can write
\begin{equation}
\tilde{g}_{t}(\tilde{z}) 
= \tilde{F}(\tilde{g}_{t+\tau}(\tilde{z})).
\label{eq:c5}
\end{equation}
The slit mapping  $\tilde{F}(\tilde{\zeta})$ can be easily computed (see Ref.~\cite{poloneses} for details), and after taking the limit $\tau\to0$, one obtains  the following Loewner equation
 \begin{equation}
 \dot{\tilde{g}}_t(\tilde{z}) =     \frac{\pi^2}{4} d(t) \frac{1-\tilde{g}_t(\tilde{z})^2}{\tilde{g}_t(\tilde{z}) -\tilde{a} },
\label{eq:f81}
\end{equation}
where
\begin{equation}
\tilde{a}(t)=\sin \left[\frac{\pi}{2}a(t)\right].
\label{eq:c3c}
\end{equation}
and
the growth factor $d(t)$ is defined by
\begin{equation}
d(t) =  \lim_{\tau\to0}\frac{h^{2}}{2 \tau}.
\label{eq:f10}
\end{equation} 
By taking the  limit $\tilde{g}_t(z)\to \tilde{a}(t)$ in Eq.~(\ref{eq:f81}) in an appropriate sense \cite{poloneses}, one finds that the driving function $a(t)$ is determined by
 \begin{equation}
 \dot{\tilde{a}} = -\frac{\pi^2}{8} d(t)\tilde{a}  .
\label{eq:f12b}
\end{equation}

The Loewner evolution in the original channel geometry can then be obtained from Eqs.~(\ref{eq:gtil}) and (\ref{eq:f81}), yielding
\begin{equation}
  \dot{g}_{t}(z) =  \frac{\pi}{2} d(t)  \frac{\cos\left[\frac{\pi}{2}g_{t}(z)\right]}{\sin\left[\frac{\pi}{2}g_{t}(z)\right]-\sin\left[\frac{\pi}{2}a(t)\right] },
\label{eq:f11}
\end{equation}
with
\begin{equation}
\dot{a}=- \frac{\pi}{4}d(t)\tan\left(\frac{\pi}{2}a\right) .
\label{eq:f12}
\end{equation}
From the boundary condition (\ref{eq:C}) it follows that $C(t)=(\pi/2)\int_0^{t}  d(t')dt'$. 
Without loss of generality we shall take $C(t)=\pi t/2$ in this section, implying that $d(t)=1$. 
Note that the Loewner equation (\ref{eq:f11})  indeed fixes the points ${z}=\pm1$ as well as  the point at infinity, in the sense of Eq.~(\ref{eq:C}).
 Hence  we shall refer to Eq.~(\ref{eq:f11})  as the tripolar Loewner equation, in analogy with the  dipolar case \cite{BB2005} where only the points $x_\pm=\pm1$ are kept fixed. The tripolar Loewner equation in the upper half-plane  given in Eq.~(\ref{eq:f81}) also fixes the points $\tilde{z}=\pm 1$ and the point at infinity. More specifically, the boundary   condition at infinity reads: $\tilde{g}_t(\tilde{z})\approx\exp[-\pi^{2}t/4]\tilde{z}$, {for} $|\tilde{z}|\to\infty$.

The Loewner equation  given in Eq.~(\ref{eq:f81}) can be readily extended to the case of multiple  curves growing simultaneously in the channel, yielding 
 \begin{equation}
 \dot{\tilde{g}}_t = \frac{\pi^2}{4} (1-\tilde{g}_t^2)\sum_{i=1}^{n}     \frac{d_{i}(t)}{\tilde{g}_t -\tilde{a}_i(t) },
\label{eq:c81}
\end{equation}
where
\begin{equation}
 \dot{\tilde{a}}_i = -\frac{\pi^2}{8}d_{i}(t) \tilde{a}_i +  \frac{\pi^2}{4} (1-\tilde{a}_i^2)\sum_{ \stackrel{j=1}{j\ne i}}^n  \frac{d_i(t)}{\tilde{a}_i -\tilde{a}_j }.
  \label{eq:121}
\end{equation}
We shall assume for simplicity  that the growth factors are all constant, i.e., $d_i(t)=d_i$, subjected to the condition $\sum_{i=1}^{n}d_{i}=1$, as follows  from our choice above for $C(t)$. Other growth models have been considered \cite{poloneses} where
the normal velocity is of the form $v_n\sim|\vec{\nabla}\phi|^{\eta}$, with  the growth factors being given by $d_i(t)=|f_{t}^{\prime\prime}(a_i(t))|^{-\eta/2-1}$, where $f_t(w)$ is the inverse of $g_t(z)$.
In this case,  the analysis  of the Loewner equation is much more difficult because of the implicit dependence of the growth factors $d_i(t)$ on the Loewner function $g_{t}(z)$. By considering the simpler (but still very interesting) case where the growth factors are constant in time, we are able both to find exact solutions for multifingers and to easily  integrate the Loewner equation (\ref{eq:c81}) on the computer.

\subsection{Exact solutions}

An exact solution for the Loewner function $g_t(z)$ for a single curve in the channel geometry can be found explicitly  for a stationary driving function, i.e., $a(t)=0$, corresponding to a finger  growing vertically along the channel centerline \cite{poloneses}. Here we  extend such stationary solutions to the case of multifingers.
For simplicity we consider only the case of two fingers, but the procedure outlined below applies, in principle, to any number of fingers (although the calculations become increasingly more difficult). 

For two fingers,   Eq.~(\ref{eq:121}) becomes
\begin{equation}
 \dot{\tilde{a}}_1 = -\frac{\pi^2}{8}\left[d_1 \tilde{a}_1 -  2 d_2 \frac{1-\tilde{a}_1^2} {\tilde{a}_1 -\tilde{a}_2 }\right],
  \label{eq:h2}
\end{equation}
\begin{equation}
 \dot{\tilde{a}}_2 = -\frac{\pi^2}{8}\left[d_2 \tilde{a}_2 - 2 d_1 \frac{1-\tilde{a}_2^2} {\tilde{a}_2 -\tilde{a}_1 }\right],
  \label{eq:h3}
\end{equation}
whose stationary solution (fixed point), $\tilde{a}_1^0$ and $\tilde{a}_2^0$, are given by
\begin{equation}
 \tilde{a}_1^0=-\left[\frac{2d_2^2(1+d_1)}{1+d_2}\right]^{1/2}, \qquad   \tilde{a}_2^0=\left[\frac{2d_1^2(1+d_2)}{1+d_1}\right]^{1/2}.
  \label{eq:h6}
\end{equation}
Note that if $d_1=d_2$, one has $\tilde{a}_1^0=-1/2$ and $\tilde{a}_2^0=1/2$, and the solution  corresponds to two identical fingers growing vertically. This symmetrical two-finger solution  is, of course, related to the single-finger solution described in Ref.~\cite{poloneses}  by a simple transformation, since in this case we can view each finger as growing along the centerline of a channel with half the width of the original channel. Thus, we are mainly interested here in asymmetrical solutions where $d_{1}\ne d_{2}$. In the remainder of this section we shall omit, for convenience of notation,  the upperscripts from $\tilde{a}_1^0$ and $\tilde{a}_2^0$. 

The Loewner equation (\ref{eq:c81}) for two fingers reads
\begin{equation}
 \dot{\tilde{g}}_t = \frac{\pi^2}{4} \frac{(1-\tilde{g}_t^2)[\tilde{g}_t-(d_1\tilde{a}_2+d_2\tilde{a}_1)]}{(\tilde{g}_t -\tilde{a}_1)(\tilde{g}_t -\tilde{a}_2)},
\label{eq:h8}
\end{equation}
where we have used the fact that $d_1+d_2=1$. This equation can be easily integrated, yielding an implicit solution for the mapping $\tilde{g}_t$ in the form
\begin{equation}
\left[\frac{\tilde{g}_t+1}{\tilde{z}+1}\right]^{\alpha_1} \left[\frac{\tilde{g}_t-1}{\tilde{z}-1}\right]^{\alpha_2} 
\left[\frac{\tilde{g}_t-(d_1\tilde{a}_2+d_2\tilde{a}_1)}{\tilde{z}-(d_1\tilde{a}_2+d_2\tilde{a}_1)}\right]^{\alpha_3}= e^{-\frac{\pi^2}{2} t},
\label{eq:h9}
\end{equation}
where
\begin{equation}
\alpha_1=\frac{(1+\tilde{a}_1)(1+\tilde{a}_2)}{1+(d_2\tilde{a}_1+d_1\tilde{a}_2)},\qquad 
\alpha_2=\frac{(1-\tilde{a}_1)(1-\tilde{a}_2)}{1-(d_2\tilde{a}_1+d_1\tilde{a}_2)},
\end{equation}
and $\alpha_3=2-\alpha_1-\alpha_2$.

The tip trajectories $\gamma_i(t)$, for $i=1,2$, can now be obtained by setting $\tilde{z}=\tilde{\gamma}_i $ and $\tilde{g}(\tilde{\gamma}_i)=\tilde{a}_i$ in Eq.~(\ref{eq:h9}), where  $\tilde{\gamma}_i=\sin(\pi\gamma_i/2)$. One then obtains
\begin{equation}
\left[\frac{\tilde{\gamma_1}+1}{\tilde{a}_1+1}\right]^{\alpha_1} \left[\frac{\tilde{\gamma_1}-1}{\tilde{a}_1-1}\right]^{\alpha_2} 
\left[\frac{\tilde{\gamma_1}-(d_1\tilde{a}_2+d_2\tilde{a}_1)}{d_1(\tilde{a}_1-\tilde{a}_2)}\right]^{\alpha_3}= e^{\frac{\pi^2}{2} t},
\label{eq:t1}
\end{equation}
\begin{equation}
\left[\frac{\tilde{\gamma_2}+1}{\tilde{a}_2+1}\right]^{\alpha_1} \left[\frac{\tilde{\gamma_2}-1}{\tilde{a}_2-1}\right]^{\alpha_2} 
\left[\frac{\tilde{\gamma_2}-(d_1\tilde{a}_2+d_2\tilde{a}_1)}{d_2(\tilde{a}_2-\tilde{a}_1)}\right]^{\alpha_3}= e^{\frac{\pi^2}{2} t},
\label{eq:t2}
\end{equation}
where the roots for $\tilde{\gamma}_1$ and $\tilde{\gamma}_2$ must be chosen in the upper half-plane.
The asymptotic trajectories in the limit $t\to\infty$ can be found by writing $\gamma_i=x_i+iy_i$ and considering the limit $y_i\to\infty$. 
After a straightforward calculation one obtains 
\begin{equation}
\gamma_1(t)\approx \frac{\alpha_1-\alpha_2-\alpha_3}{2}+i\frac{\pi}{2}t, \qquad \gamma_2(t)\approx \frac{\alpha_1-\alpha_2+\alpha_3}{2}+i\frac{\pi}{2}t, \qquad t\to\infty.
\label{eq:f16}
\end{equation}
We thus see that for large times the two fingers approach the vertical axes 
\begin{equation}
x_{1,2}=(\alpha_1-\alpha_2\mp\alpha_3)/2,
\end{equation}
respectively, and both of them move asymptotically with the constant speed $v=\pi/2$.  
 
\begin{figure}[t]
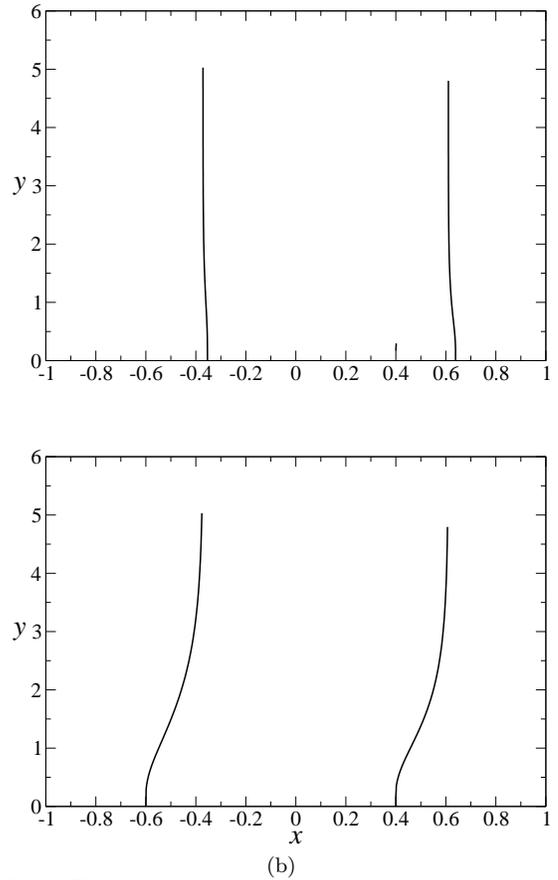

\begin{center}
\subfigure[\label{fig:2a}]{\includegraphics[width=0.4\textwidth]{fig2a.eps}}\hspace{10mm}
\subfigure[\label{fig:2b}]{\includegraphics[width=0.4\textwidth]{fig2b.eps}}
\end{center}
\caption{Evolution of two fingers with $d_1=2/3$ and $d_2=1/3$. The initial conditions are: $a_1(0)=a_1^0=-0.3534$ and $a_2(0)=a_2^0=0.6387$ in (a), and $a_1(0)=-0.6$ and $a_2(0)=0.6$ in (b). In both cases the trajectories approach the axes $x_1=-0.3712$ and $x_2=0.6099$.}
\label{fig:2}
\end{figure}

An example of such a solution is shown in Fig.~\ref{fig:2a}, where we used  $d_1=2/3$ and $d_2=1/3$, which from (\ref{eq:h6}) implies  
$a_1^0=-0.3534$ and $a_2^0=0.6387$.  In this case, the fingers should approach the vertical axes $x_1=-0.3712$ and $x_2=0.6099$, as predicted by (\ref{eq:f16}), and this is indeed observed in  the figure. Note that this  asymptotic behavior holds for any initial conditions $a_1(0)$ and $a_2(0)$, on account of the fact that the fixed point given in (\ref{eq:h6}) is stable. This is illustrated in Fig.~\ref{fig:2b}  where we show a solution  with initial conditions   $a_1(0)=-0.6$ and $a_2(0)=0.4$. 
To generate the curves shown in Fig.~\ref{fig:2} we integrated the Loewner equation given in Eq.~(\ref{eq:c81}) using the numerical scheme described in \cite{us_pre}.

\subsection{Tripolar Loewner Chains}

In the spirit of chordal Loewner chains \cite{BB}, it is clear that the tripolar Loewner equation (\ref{eq:c81}) can be extended to describe the growth of more general domains in the channel geometry, not restricted to a curve or a set of curves. The corresponding tripolar Loewner chain thus reads 
\begin{equation}
\dot{g}_t(z)=\left(1-g_t(z)^2\right)\int_{-1}^1\frac{\rho_t(x)  dx}{g_{t}(z)-x}\, ,
\label{eq:Lc}
\end{equation}
where the density of singularities $\rho_t(x)$ can be viewed as a measure of the growth rate at a point $z$ at the boundary of the growing set that is the preimage of $x$ under $g_{t}(z)$. Since the shape of the growing domain is fully encoded in the map $g_{t}(z)$, equation (\ref{eq:Lc}) specifies the growth model once the density $\rho_t(x)$ is known. As discussed elsewhere \cite{us_pre},  the formalism of Loewner chains is of limited practical use, except when the density is a sum of Dirac $\delta$-functions, in which case we are back to multiple growing curves. In this context, it is important to consider more explicit  models to describe the growth of a domain bounded by an interface. One such model was first introduced in Ref.~\cite{us_pre} for the upper half-plane, and in the next section it is extended to the channel geometry.

\section{Loewner Evolutions  for Interfaces in the Channel}
\label{sec:3}

\begin{figure}[t]
\centering
\includegraphics[width=0.45\textwidth]{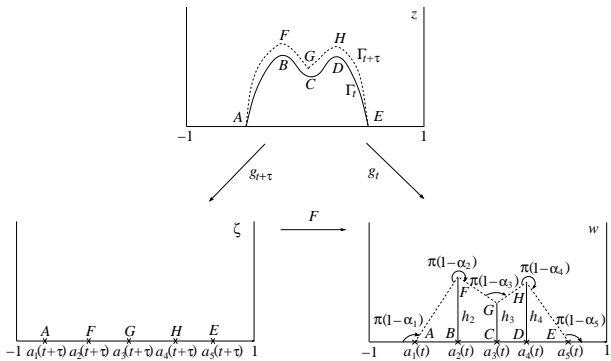}
\caption{The physical and mathematical planes for a growing interface in the channel geometry. The mapping $g_{t}(z)$ maps the interface $\Gamma_t$ to a segment on the real axis, while the new interface  $\Gamma_{t+\tau}$ is mapped
to a tent-like shape.}
\label{fig:3}
\end{figure}

\subsection{Tripolar Loewner Equation for Interfaces}

Here we consider the problem of an interface starting from a segment $[z_1,z_2]\subset(-1,1)$ on the real axis in the $z$-plane and growing into the upper half-channel,  as indicated in Fig.~\ref{fig:3}.  
In our growth model we assume, for simplicity, that the growing interface has certain
special points, referred to as \emph{tips}  and \emph{troughs}, where the growth rate is  a local maximum and a local minimum, respectively, while the interface endpoints $z_1$ and $z_2$ remain fixed.
Let us denote by $\Gamma_t$ the interface at time $t$  and by $K_t$ the growing region delimited by $\Gamma_t$ and the real axis. It is  assumed that the curve $\Gamma_t$ is simple so that the physical domain $\mathbb{P}\backslash K_t$ is simply connected, hence $K_t$ is a hull. As before, the Loewner function $g_t(z)$ maps the physical  domain  in the $z$-plane onto the upper half-channel $\mathbb{P}$  in  the mathematical plane $w$-plane, that is,
\begin{equation*}
 g_t:\mathbb{P}\backslash K_t \rightarrow \mathbb{P}  ,
\end{equation*}
Under  the action of  $g_t(z)$, the interface $\Gamma_t$ is mapped  to an interval on the real axis in the $w$-plane, with the images of the tips, troughs and end points being denoted by $a_i(t)$, $i=1,...,N$, where $N$ is the total number of such special points; see Fig.~\ref{fig:3}. 
The mapping function  $g_t(z)$ is again required to satisfy satisfy the hydrodynamic condition (\ref{eq:C}) and the initial condition (\ref{eq:f0}). Here, however, we shall not fix the function $C(t)$ {\it a priori}.

\begin{figure}[t]
\begin{center}
\includegraphics[width=0.45\textwidth]{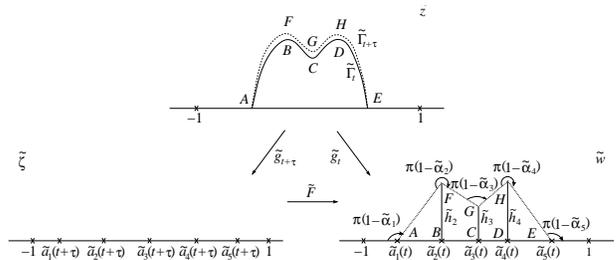}
\end{center}
\caption{The transformed physical and the mathematical planes  for a growing interface. The `tilded' planes are obtained from the original domains shown in Fig.~\ref{fig:3} by applying the transformation $\phi(z)= \sin\left(\frac{\pi}{2} z\right)$; see text.}
\label{fig:3b}
\end{figure}

The growth dynamics in our model is specified by requiring that tips and troughs  grow along gradient lines in such a way that  the interface $\Gamma_{t+\tau}$ at time $t+\tau$, for infinitesimal $\tau$,  is mapped under $g_t(z)$ to a polygonal  curve,  as shown in  Fig.~\ref{fig:3}. 
The exterior angle at the $i$-th vertex of our polygonal curve is denoted by $\pi(1-\alpha_{i})$, with the convention that if the angle is greater than $\pi$ the corresponding parameter $\alpha_{i}$ is negative.  In other words, the parameter $\alpha_{i}$'s are negative for tips  and positive for troughs and end points. From a trivial trigonometric relation it follows that  
\begin{equation}
\sum_{i=1}^{N}\alpha_i=0.
\label{eq:suma}
\end{equation}
If we denote by ${h}_i$ the heigth of the $i$th vertex of the polygonal curve  in the $w$-plane,
then as $\tau \to 0$ the parameters $h_i$ and $\alpha_{i}$  all go to zero, and in this limit one finds the following relation: 
 \begin{equation}
\sum_{i=1}^{N}a_{i}\alpha_i =0.
\label{eq:suma2}
\end{equation}

As before,  the mathematical plane at time $t+\tau$ is represented by the complex $\zeta$-plane, so that under the mapping $\zeta=g_{t+\tau}(z)$,  the interface tips, troughs, and end points  at time $t+\tau$  are mapped to the points $\zeta=a_i(t+\tau)$, as indicated in Fig.~\ref{fig:3}. To derive the tripolar Loewner equation for the growing interface we first need to obtain the mapping $w=F(\zeta)$, so that we can write $g_t(z)=F(g_{t+\tau}(z))$, and then take the limit $\tau\to0$. As discussed in Sec.~\ref{sec:2}, it is more convenient  however to work with the tripolar Loewner equation in the upper half-plane by applying the transformations given in Eq.~(\ref{eq:sin}). The corresponding  transformed domains are shown in Fig.~\ref{fig:3b}. In the $\tilde{w}$-plane, we obtain (in the limit of $\tau\to0$) a polygonal curve  whose vertices are located at the points $\tilde{a}_i+i\tilde{h}_i$, where $\tilde{a}_i=\sin(\pi a_i/2)$ and  the heigth $\tilde{h}_i$   in the $\tilde{w}$-plane is given 
by
\begin{equation}
 \tilde{h}_{i} = \frac{\pi}{2} \cos \big( \frac{\pi}{2} a_i \big) h_i,
 \label{eq:hit2}
\end{equation}
and the exterior angle at the $i$th vertex is denoted by  $\pi(1-\tilde{\alpha}_{i})$. Relations analogous to (\ref{eq:suma}) and (\ref{eq:suma2}) are obviously valid for  $\tilde{\alpha}_i$ and $\tilde{a}_i$. 

Next consider the mapping, $\tilde{w}=\tilde{F}(\tilde{\zeta})$, from the upper half-$\tilde{\zeta}$-plane onto the  upper half-$\tilde{w}$-plane
with a polygonal cutout region; see Fig.~\ref{fig:3b}. Since the domain in the $\tilde{w}$-plane can be viewed as a degenerate polygon, the mapping function $\tilde{F}$ can be easily obtained from the Schwarz-Christoffel transformation \cite{CKP}, which yields
\begin{equation}
\tilde{g}_{t}=\tilde{F}(\tilde{g}_{t+\tau})= K \varphi(\tilde{g}_{t+\tau}) + C,
 \label{eq:gtB}
\end{equation}
where 
\begin{equation}
\varphi(\tilde{\zeta}) = \int^{\tilde{\zeta}} \prod_{i=1}^N{[z-\tilde{a}_i(t+\tau)]^{-\tilde{\alpha}_i}}\,dz .
 \label{eq:varphi}
\end{equation}
The constants $K$ and $C$ are determined so as to guarantee that $\tilde{F}(\pm1)=\pm1$,  and hence we can rewrite Eq.~(\ref{eq:gtB})  as
\begin{equation}
 \tilde{g}_t=\frac{2\varphi(\tilde{g}_{t+\tau})-[\varphi(1)+\varphi(-1)]}{\varphi(1)-\varphi(-1)}.
 \label{eq:1d}
 \end{equation}

The integral  in Eq.~(\ref{eq:varphi}) cannot, in general, be performed explicitly. It is thus convenient to expand the integrand in powers of  the infinitesimal quantities $\tilde{\alpha}_i$ and then proceed with the integration term by term. Doing this up to the first order in the
 $\tilde{\alpha}_i$'s,  one finds
\begin{equation}
\varphi(\zeta)= \zeta - \sum_{i=1}^N \tilde{\alpha}_i(t)[\zeta -\tilde{a}_i(t+\tau) ]\ln[\zeta -\tilde{a}_i(t+\tau)]  .
\label{eq:1c2}
\end{equation}
If one now inserts Eq.~(\ref{eq:1c2}) into Eq.~(\ref{eq:1d}), expand the resulting expression up to first order in $\tau$, 
and then take $\tau \to 0$,   one obtains the following Loewner equation
\begin{equation}
 \dot{\tilde{g}}_t(z)= \sum_{i=1}^N \tilde{d}_i(t)\left\{[\tilde{g}_{t} -\tilde{a}_i(t) ]\ln[\tilde{g}_{t} -\tilde{a}_i(t)] - A_i(t) \tilde{g}_t+B_i(t)\right\} ,
\label{eq:L}
\end{equation}
with
 \begin{equation}
A_i(t)= \frac{1}{2} \left\{[1 +\tilde{a}_i(t) ]\ln[1 +\tilde{a}_i(t)] +[1 -\tilde{a}_i(t) ]\ln[1 -\tilde{a}_i(t)]\right\},
\label{eq:A}
\end{equation}
 \begin{equation}
B_i(t)= \frac{1}{2} \left\{[1 +\tilde{a}_i(t) ]\ln[1 +\tilde{a}_i(t)] -[1 -\tilde{a}_i(t) ]\ln[1 -\tilde{a}_i(t)]\right\},
\label{eq:B}
\end{equation}
where
the `tilded' growth factors $\tilde{d}_i(t)$ are defined by
\begin{equation}
\tilde{d}_i(t)= \lim_{\tau \to 0}\frac{\tilde{\alpha}_i}{\tau}.
\label{eq:dit}
\end{equation}

The time evolution of the driving functions $\tilde{a}_i$ is determined by requiring that the points $\tilde{a}_i(t+\tau)$ be taken under the mapping $F$ to the points $\tilde{a}_i(t)$.  In other words, we impose that
\begin{equation}
\tilde{a}_{i}(t)= F(\tilde{a}_{i}(t+\tau)).
 \label{eq:atC}
\end{equation}
Applying the procedure used above to derive the Loewner equation Eq.~(\ref{eq:L}), it  immediately follows that the  differential equations governing the dynamics of the driving functions $\tilde{a}_i$ can be obtained by simply setting $\tilde{g}_t=\tilde{a}_i$ in  Eq.~(\ref{eq:L}), which yields
\begin{equation}
 \dot{\tilde{a}}_i= 
 \sum^N_{j=1} \tilde{d}_j(t)\left\{[\tilde{a}_{i} -\tilde{a}_j(t) ]\ln\left|\tilde{a}_{i} -\tilde{a}_j(t)\right| - A_j(t) \tilde{a}_i+B_j(t)\right\} .
\label{eq:dotai}
\end{equation}

\begin{figure}[b]
\vspace{1cm}
\begin{center}
\includegraphics[width=0.4\textwidth]{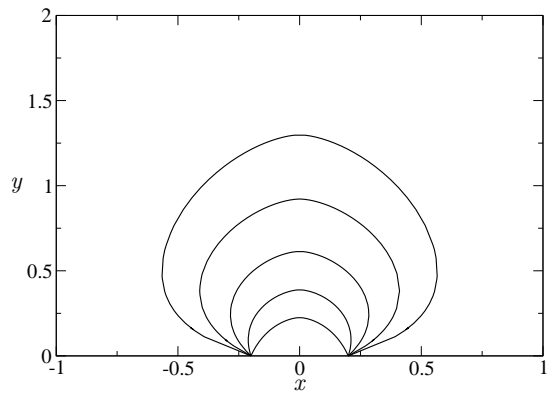}
\end{center}
\caption{Loewner evolution for a symmetric interface growing in the channel geometry. The solid curves show the interface  at various times $t$, starting from $t=0.5$ up to $t=2.0$ with  a time separation of  $\Delta t =0.3$ between successive curves.}
\label{fig:5}
\end{figure}

The tilded growth factors $\tilde{d}_i$ defined in Eq.~(\ref{eq:dit}) can be related to the growth factors, ${d}_i(t)= \lim_{\tau \to 0}\left({{\alpha}_i}/{\tau}\right)$, defined for the original channel geometry 
by virtue of trivial trigonometric relations. After some calculation, one finds
\begin{equation}
\tilde{d}_i = \frac{{C}_{i+1}-{C}_{i}}{\tilde{a}_{i+1} - \tilde{a}_{i}} + \frac{{C}_{i-1}-{C}_{i}}{\tilde{a}_{i} - \tilde{a}_{i-1}}  ,  \qquad 2 \le i \le N-1,
\label{eq:dit2}
\end{equation}
where
\begin{equation}
 C_i = \frac{\pi}{2}\cos\left(\frac{\pi}{2}a_i\right) \sum_{j=1}^{i-1} d_{i-j} (a_i - a_{i-j}). 
 \label{eq:Ci}
\end{equation}
From the Eqs.~(\ref{eq:suma}) and (\ref{eq:suma2}) it also follows that the growth factors $d_i$ satisfy the following relations
\begin{equation}
\sum_{i=1}^{N}d_i=0,
\label{eq:sumd}
\end{equation}
 \begin{equation}
\sum_{i=1}^{N}a_{i}d_i =0.
\label{eq:sumd2}
\end{equation}
In view of the relations above, we can express the parameters $d_{1}(t)$ and $d_{N}(t)$ associated with the end points  in terms of the other growing factors $d_{i}(t)$, so that our growth model is completely specified  by prescribing the functions $d_{i}(t)$ at the tips and troughs of the interface.  Thus, once the growth factors $d_i$ are given, one can compute the tilded growth factors $\tilde{d}_i$ from Eq.~(\ref{eq:dit2}), proceed with the integration of the Loewner equation given in Eq.~(\ref{eq:L}), and then invert the transformation  (\ref{eq:gtil}) to obtain the Loewner mapping $g_t(z)$.  Next we discuss some examples of   interface growth described by the above model.

\subsection{Examples}

\subsubsection{Single Tip}

For an interface with a single tip (i.e., $N=3$), the factors $d_1(t)$ and $d_3(t)$ can be written in terms of $d_2(t)$, so that we can set $d_2=-1$ without loss of generality, since this amount to a mere  rescaling of the time coordinate.  Let us consider first the case in which the interface is  symmetric with respect to the channel centerline:  $a_2(t)=0$ and $a_3(t)=-a_1(t)\equiv a(t)$. This imply, in particular, that $\tilde{d}_1=\tilde{d}_3=-\tilde{d}_2/2$. Introducing the parameter $\tilde{d}(t)=| \tilde{d}_2(t)|$, the Loewner equation (\ref{eq:L}) then becomes
\begin{widetext}
\begin{equation}
 \dot{\tilde{g}}_t(z)=  \frac{1}{2}\tilde{d}(t)\left\{[\tilde{g}_{t} +\tilde{a}(t) ]\ln[\tilde{g}_{t} +\tilde{a}(t)]+[\tilde{g}_{t} -\tilde{a}(t) ]\ln[\tilde{g}_{t} -\tilde{a}(t)] -2\tilde{g}_{t}\ln\tilde{g}_{t}- A_s(t) \tilde{g}_t\right\},
\label{eq:Ls}
\end{equation}
\end{widetext}
where
 \begin{equation}
A_s(t)=   [1 +\tilde{a}(t) ]\ln[1 +\tilde{a}(t)] +[1 -\tilde{a}(t) ]\ln[1 -\tilde{a}(t)],
\label{eq:Ap}
\end{equation}
with the dynamics for $\tilde{a}(t)$ being given by
\begin{equation}
 \dot{\tilde{a}}(t)= \tilde{d}(t) \left( 2\ln 2- A_s \right)\tilde{a},
\label{eq:ast}
\end{equation}
In this case, one can show from Eq.~(\ref{eq:dit2}) that $\tilde{d}(t)$ becomes
\begin{equation}
\tilde{d}(t)=\frac{\pi}{2} \tan\left[\frac{\pi}{2}a(t)\right]=\frac{\pi}{2}\frac{\tilde{a}(t)}{\sqrt{1-\tilde{a}^2(t)}},
\end{equation}
where we have used the fact that $|d_2|=1$.

\begin{figure}[b]
\vspace{1cm}
\begin{center}
{\includegraphics[width=0.4\textwidth]{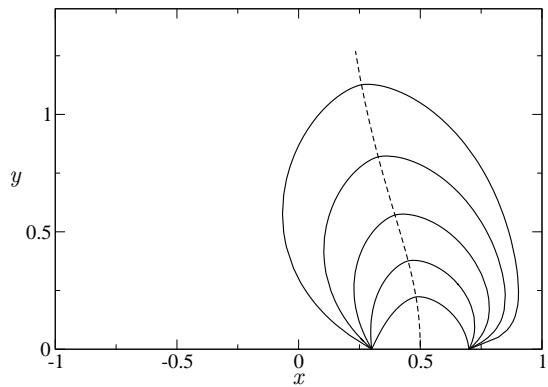}}
\end{center}
\caption{Loewner evolution for an  asymmetric interface growing in the channel geometry. The solid curves show the interface  at several times  from $t=0.5$ up to $t=1.7$, with   $\Delta t =0.3$ between successive curves, while the dashed curve  represents the path traced by the tip.}
\label{fig:6}
\end{figure}

 An example of a symmetric interface in a channel  is shown in Fig.~\ref{fig:5}.   In this figure, the solid curves represent the interface  at various times $t$, starting from $t=0.5$ up to $t=2.0$ with  a time separation of  $\Delta t =0.3$ between successive curves.  To generate the curves shown in Fig.~\ref{fig:5}  we integrated the Loewner equation (\ref{eq:Ls}) backwards in time from   `terminal conditions' $\tilde{g}_t=\tilde{w}$, for $\tilde{w}\in [-\tilde{a}(t),\tilde{a}(t)]$, to obtain the respective initial values $\tilde{g}_0$ from which we determine the corresponding points $z$ on the interface: $z=({2}/{\pi})\arcsin\left(\tilde{g}_0\right)$.  One sees from the figure that as time proceeds the interface expands and tends to occupy the entire channel. It is worth mentioning, however, that it is hard to go past the latest time shown in Fig.~\ref{fig:5} because the function $a(t)$ becomes very close to unity, rendering the numerical integration  of the Loewner equation very difficult after this point.

In the case of an asymmetric interface, we need to integrate the full Loewner equation given in Eq.~(\ref{eq:L}) with $N=3$.
In   Fig.~\ref{fig:6} we show a solution where the solid curves represent the interface  at various times $t$ (see caption for details), while the dashed curve  indicates the path traced by the tip $\gamma_{t}=g^{-1}_{t}(a_{2}(t))$. As seen  in Fig.~\ref{fig:6}, the tip is ``repelled" by its image with respect to the closest channel wall, so that  an asymptotically symmetrical  shape is expected for sufficiently long time. (It is difficult, however, to integrate the Loewner equation for larger times  for reasons already mentioned.)

\begin{figure}[b]
\vspace{1cm}
\centering
\includegraphics[width=0.4\textwidth]{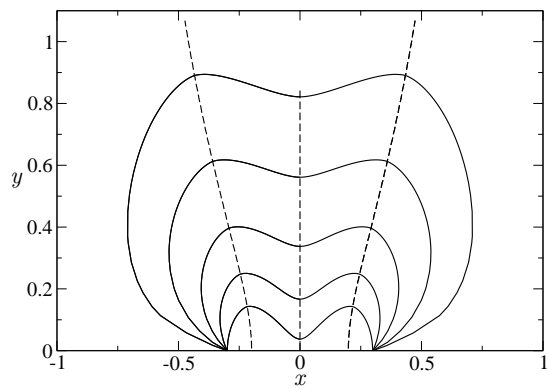}
\caption{Symmetrical growing interface with two tips in the channel for $d_{2}=d_{4}=-2$ and $d_3=1.5$.
The solid curves represent the interface at times from $t=0.5$ to $t=1.7$,
with $\Delta t = 0.3$ between successive curves, whereas the dashed lines indicate the trajectories of  the tips and the trough.}
\label{fig:8}
\end{figure}

\subsubsection{Multiple Tips}

\label{sec:3b}

\begin{figure}[t]
\centering
\includegraphics[width=0.4\textwidth]{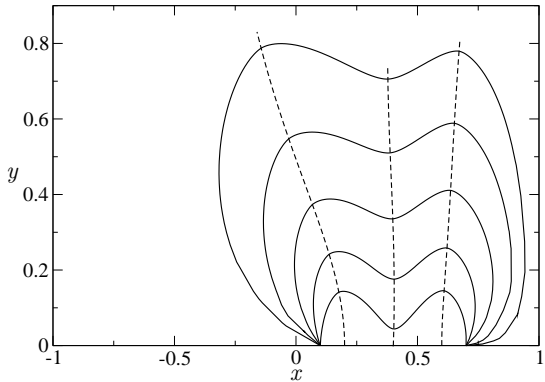}
\caption{Asymmetrical growing interface with two tips  in the channel. The growth factors  are the same as in Fig.~\ref{fig:8} but  the interface is initially located off center.}
\label{fig:11}
\end{figure}

Let us now illustrate some of the patterns that emerge from our growth model for interfaces with two tips and one trough (i.e., $N=5$), in which case tip competition can arise. Here we shall consider  for simplicity only cases in which  $d_{i}(t)=d_{i}={\rm const.}$
We expect nevertheless that the qualitative behavior  seen in this simpler situation  (see below) should also be valid in more general cases.
As a first example, we show in Fig.~\ref{fig:8}  a symmetrical growing interface, where the dashed lines represent the trajectories of the tips and the trough. In this figure,  we started with symmetrical initial conditions, namely, $a_{5}(0)=-a_{1}(0)=0.3$, $a_{4}(0)=-a_{2}(0)=0.2$, and $a_{3}(0)=0$, and chose the growth factors of the two tips to be the same, $d_{2}=d_{4}=-1$, so as to preserve the initial symmetry, with the trough growth factor being $d_3(t)=0.6$.

If we break the symmetry, either by choosing an asymmetric initial condition or using different growth factors for the tips, then inevitably of the tips will grow faster and ``screen'' the other tip. 
For instance, in  Fig.~\ref{fig:11} we show a situation where  the growth  factors  are the same as in  Fig.~\ref{fig:8} but the initial segment from which the interface grows is off-center, namely, $a_1(0)=0.2$, $a_2(0)=0.3$, $a_3(0)=0.5$, $a_4(0)=0.7$, and $a_5(0)=0.8$.  In this case, the tip  closest to the right channel wall grows slower and falls  behind the other tip. Here, however, the screening is partial in the sense that the slower tip continues to grow but with a velocity that is a fraction of that of the faster tip. (Total screening, whereby the slowest tip eventually stops growing altogether, can be observed if the faster tip has a sufficiently large growth factor.)

\subsubsection{Multiples Interfaces}

\begin{figure}[t]
\begin{center}
\includegraphics[width=0.4\textwidth]{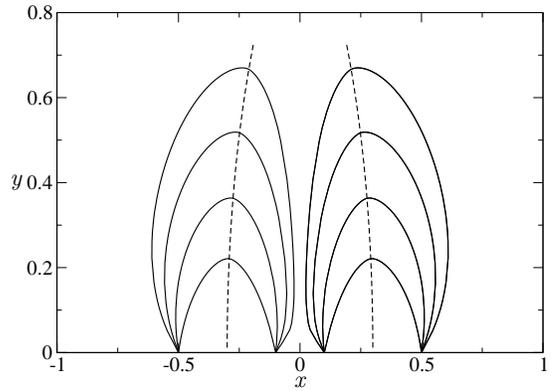}
\end{center}
\caption{Loewner evolution for two symmetrical surfaces growing in the the channel geometry for $d(t)=1$ for each interface. The interfaces start respectively at $[-3,-1]$ and $[1,3]$. The different curves represent the interfaces at time intervals of $\Delta t = 0.3$.}
\label{fig:12}
\end{figure}

The generic Loewner equation given in Eq.~(\ref{eq:L}) can also describe the problem of multiple  growing interfaces in the channel geometry.
 In this case each interface $\Gamma^i_t$, for $i=1,...,n$, where $n$ is the number of distinct interfaces,  will be mapped under $g_t(z)$ to a corresponding interval on the real axis in the $w$-plane. Similarly, each advanced interface  $\Gamma^i_{t+\tau}$ is mapped by $g_t(z)$ to a polygonal curve in the $w$-plane. One can readily convince oneself that the generic Loewner evolution defined by (\ref{eq:L}) and (\ref{eq:dotai}) applies to this case as well, where  $N$ is now the total number of vertices corresponding to the sum of the number of vertices for each interface. Note that for each interface we have conditions analogous to (\ref{eq:sumd}) and (\ref{eq:sumd2}), but involving only the growth factors $d_i$ of the respective interface.

\begin{figure}[b]
\vspace{0.5cm}
\begin{center}
\includegraphics[width=0.4\textwidth]{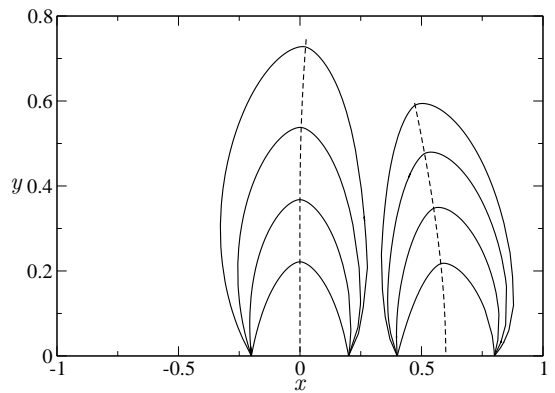}
\end{center}
\caption{ Loewner evolution for two asymmetric growing fingers. Here 
the growth factors are the same as in
Fig.~\ref{fig:12} but the initial segments from which the interface starts to grow were shifted to the right.}
\label{fig:13}
\end{figure}

In Fig.~\ref{fig:12} we show numerical solutions for two growing symmetrical interfaces, with one tip each, where the interfaces are the images of one another with respect to the $y$ axis. 
Here the interfaces start to grow from the intervals $[-3,-1]$ and $[1,3]$, respectively, with the corresponding tips starting at symmetrical points $a_4(0)=-a_2(0)=2$. (The interfaces have the same growing factors, meaning that $d_2=d_4=-1$.) 
Notice that, as time goes by, the inner sides of the two interfaces  move towards one another leaving a narrow channel between them. For sufficiently large time, the width of such a channel becomes infinitesimally small so that for all practical purposes the resulting evolution will look like a single,  symmetrical interface.

If  the symmetry between the two interfaces is broken, then one of them will eventually move ahead of the other one. For instance, in Fig.~\ref{fig:13} we show  the case in which the growth factors are the same as in Fig.~\ref{fig:12} but the initial positions of the interfaces are no longer symmetric with respect to the channel centerline. In this case,  the interface closer to the centerline is the ``winner," in the sense that its tips moves ahead of the tip of the second interface. 
This competition between the two growing interfaces resembles the so-called ``shadowing effect" in fingering phenomena, whereby the longer fingers grow faster and hinder the growth of the shorter ones in their vicinities.  

\section{Conclusions}
\label{sec:4}

We have discussed a class of growth models in the channel geometry where the growth dynamics is described  in terms of tripolar Loewner evolutions. In the tripolar geometry, the Loewner mapping fixes the points $z=\pm 1$ and the point at infinity. The growing domain starts  from a point (in the case of a curve) or from an interval (in the case of a domain bounded by an interface) on the real axis, ending at infinity (for infinite time).  In particular, a novel class of exact solutions of the tripolar Loewner equation for multiple curves was obtained for the case of stationary driving functions. As for the more difficult problem of interface growth in the channel geometry, we introduced a generalized tripolar Loewner equation and presented several numerical solutions, including the case of an interface with multiple tips as well as the case of multiple interfaces.  It was argued that the behavior seen in the model is reminiscent of the phenomenon of  tip/finger competition observed in fingering experiments.

A natural extension of the work presented here would be to consider the problem of tripolar stochastic  Loewner equation (SLE) and investigate its connection with statistical mechanics models and conformal field theory. Although stochastic growth processes were not considered in this paper,  it is nonetheless hoped that a better understanding of deterministic Loewner evolutions in the channel geometry, which was the main thrust of our study, may be of some help in tackling the more difficult problem  of tripolar SLEs. Another possible direction for future work would be to consider more general geometries, such as ``quadripolar'' Loewner evolutions where four points are kept  fixed. This geometry would be of interest in connection, for instance, with growth processes in doubly periodic domains.

\begin{acknowledgments}
This work was supported in part by the Brazilian agencies FINEP, CNPq, and FACEPE and by the special programs PRONEX and CT-PETRO. 
\end{acknowledgments}



\vspace{-0.5cm}
\begin{thebibliography}{}
%
%

%
\bibitem{loewner} L\"owner, K.: Untersuchungen \"uber schlichte konforme Abbildungen des Einheitskreises. I. Math.~Ann.~{\bf 89}, 103--121 (1923).

\bibitem{review3} Lawler, G.: Introduction to the stochastic Loewner evolution. In: Random Walks and Geometry, pp. 261--294. Walter de Gruyter, Berlin (2004).


\bibitem{BB2005} Bauer, M., Bernard, D.: Dipolar stochastic Loewner evolutions, {J.~Stat.~Mech.}  P03001 (2005). {ArXiv:math-ph/0411038}.

\bibitem{poloneses} Gubiec, T.,  Szymczak, P.: Fingered growth in the channel geometry: A Loewner-equation approach. Phys.~Rev.~ E {\bf 77}, 041602-1--041602-12 (2008).

\bibitem{schramm} Schramm, O.: Scaling limits of loop-erased random walks and uniform spanning trees. Isr.~J.~Math.~{\bf118}, 221--288 (2000).

\bibitem{review1} Gruzberg, I.~A., Kadanoff, L.~P.:
The Loewner equation: Maps and shapes. J.~Stat.~Phys.~{\bf 114}, 1183--1198 (2004). 

\bibitem{review2} Kager, W., Nienhuis, B.: A guide to stochastic Loewner evolution and its applications. J.~Stat.~Phys.~{\bf 115}, 
1149--1229 (2004).

\bibitem{cardy} Cardy, J.: SLE for theoretical physicists. Ann.~Phys.~{\bf 318}, 81--118 (2005).

\bibitem{BB} Bauer, M., Bernard, D.: 2D growth processes: SLE and Loewner chains. Phys.~Rep.~{\bf 432}, 115--221 (2006).

\bibitem{kadanoff_jsp} Kager, W., Nienhuis, B., Kadanoff, L.~P.: Exact solutions for Loewner evolutions. J.~Stat.~Phys.~{\bf 115}, 805--822 (2004).

\bibitem{selander} Selander, G: Two deterministic growth models related to diffusionn limited aggregation. Ph.D. thesis, Royal Institute of Technology, Stockholm (1999).

\bibitem{makarov} Carleson, L., Makarov, N.: Laplacian path models.  J.~Anal.~Math.~{\bf 87}, 103--150 (2002). 

\bibitem{zabrodin2} Zabrodin, A.: Growth processes related to the dispersionless Lax equations. Physica D {\bf 235}, 101--108 (2007).

\bibitem{combustion} Zik, O., Moses, E.: Fingering instability in combustion: An extended view. Phys.~Rev.~E {\bf 60}, 518--531 (1999).

\bibitem{us_pre} Dur\'an, M.~A., Vasconcelos, G.~L.: Interface growth in two dimensions: A Loewner-equation approach. Phys.~Rev.~E {\bf 82}, 031601-1--03161-8 (2010).

\bibitem{zabrodin1} Zabrodin, A.: Growth of fat slits and dispersionless KP hierarchy. J. Phys. A: Math. Theor.  {\bf 42}, 085206 (2009).

\bibitem{pelce} P.~Pelc\'e, {\it Dynamics of curved fronts}  (Academic Press, San Diego, 1988).

\bibitem{CKP} G.~F.~Carrier, M.~Krook, and C.~E.~Pearson, {\it Functions of a complex variable: theory and technique} (Hod Books, Ithaca, 1983).

\end{thebibliography}
\end{document}